\documentclass[aps,amssymb,amsmath,prl,reprint,noshowpacs,superscriptaddress]{revtex4-1}
\usepackage[english]{babel}
\usepackage[utf8]{inputenc}
\usepackage{amsthm}
\usepackage{mathtools}
\usepackage{physics}
\usepackage{xcolor}
\usepackage{graphicx}
\usepackage[left=23mm,right=13mm,top=35mm,columnsep=15pt]{geometry} 
\usepackage{adjustbox}
\usepackage{placeins}
\usepackage[T1]{fontenc}
\usepackage{lipsum}
\usepackage{csquotes}
\usepackage{siunitx}
\usepackage{natbib}
\usepackage{hyperref}
\usepackage{appendix}
\hypersetup{
	colorlinks   = true, 
	urlcolor     = blue, 
	linkcolor    = blue,
	citecolor   = blue 
}
\usepackage{color}
\usepackage{lipsum}
\usepackage{comment}
\usepackage{scalefnt} 

\begin{document}

\scalefont{1.0}

\title{Observation of radiation torque shot noise on an optically levitated nanodumbbell}

\author{Fons~\surname{van der Laan}}
\email[Correspondence email address: ]{vfons@ethz.ch}
\affiliation{Photonics Laboratory, ETH Z{\"u}rich, 8093 Z\"urich, Switzerland}
\author{Ren{\'e}~\surname{Reimann}}
\affiliation{Photonics Laboratory, ETH Z{\"u}rich, 8093 Z\"urich, Switzerland}
\affiliation{Quantum Sensing Laboratory, Quantum Research Centre, Technology Innovation Institute, Abu Dhabi, UAE}
\author{Felix~\surname{Tebbenjohanns}}
\affiliation{Photonics Laboratory, ETH Z{\"u}rich, 8093 Z\"urich, Switzerland}
\author{Jayadev~\surname{Vijayan}}
\affiliation{Photonics Laboratory, ETH Z{\"u}rich, 8093 Z\"urich, Switzerland}
\author{Lukas~\surname{Novotny}}
\affiliation{Photonics Laboratory, ETH Z{\"u}rich, 8093 Z\"urich, Switzerland}
\author{Martin~\surname{Frimmer}}
\affiliation{Photonics Laboratory, ETH Z{\"u}rich, 8093 Z\"urich, Switzerland}
\homepage[]{https://photonics.ethz.ch}

\date{\today} % Leave empty to omit a date

\begin{abstract}
According to quantum theory, measurement and backaction are inextricably linked. In optical position measurements, this backaction is known as radiation pressure shot noise. In analogy, a measurement of the orientation of a mechanical rotor must disturb its angular momentum by radiation torque shot noise. In this work, we observe the shot-noise torque fluctuations arising in a measurement of the angular orientation of an optically levitated nanodumbbell. 
We feedback cool the dumbbell's rotational motion and investigate its reheating behavior when released from feedback. In high vacuum, the heating rate due to radiation torque shot noise dominates over the thermal and technical heating rates in the system.

\end{abstract}

%\keywords{first keyword, second keyword, third keyword}

\maketitle

\paragraph{Introduction.---} \label{sec:introduction}
Harnessing light to measure and control mechanical motion is the central theme of optomechanics~\cite{bowen2015quantum,aspelmeyer2014cavity}.
At the heart of the paradigmatic optomechanical system is a light field interacting with a mechanical degree of freedom coupled to a thermal bath. 
The light field interrogates the mechanical motion, and, in accordance with the Heisenberg uncertainty relation, gives rise to a backaction on the mechanics~\cite{Braginski1968}. 
A pivotal step in the development of any quantum-optomechanical system is to boost the coupling between the mechanics and the light field sufficiently to overcome the interaction with the thermal bath.
In this regime, the exquisite control researchers have gained over the quantum states of light can be exploited, to perform measurements at, and even below, the standard quantum limit~\cite{Schreppler2014,Mason2019}. 
Furthermore, this regime allows for measurement-based feedback control of the mechanics outside the bounds of classical physics~\cite{Sudhir2017,Rossi2018}, a prerequisite to embark on the quest to engineer massive objects into macroscopic quantum states~\cite{Romero-Isart2011,Romero-Isart2017}.

For translational motion, the hallmark signature of the quantum nature of light dominating the dynamics of the mechanics has been the observation of radiation pressure shot noise~\cite{Purdy2013,Jain2016}.
These pressure fluctuations can be explained by viewing the light field as a stream of discrete, mutually independent photons, each carrying a linear momentum proportional to $\hbar$~\cite{Lebedev1901}. 
The statistics of this momentum transfer leads to shot-noise fluctuations of the radiation pressure. 
In quantum theory, these fluctuations arise due to an interference of the deterministic measurement field with the vacuum fluctuations~\cite{Caves1980}. 

In recent years, rotational motion has attracted increasing attention in optomechanics~\cite{Kim2016,Delord2020}. 
A torsional rotor with a linear restoring force (termed librator) resembles a harmonic oscillator. 
The application of techniques developed to control translational motion thus offers promise to turn librational degrees of freedom into a quantum resource for optomechanics. 
%On the other hand, free rotors offer particularly rich physics due to the cyclic nature of their angular coordinate~\cite{Shi2016}. 
Prominent examples are quantum revivals~\cite{Stickler2016,Stickler2018,Stickler2018rotationquantum}, which may offer an alternative route to explore quantum mechanics at a macroscopic scale, as well as quantum friction at extreme rotation frequencies~\cite{Manjavacas2010,Zhao2012,Xu2017,Manjavacas2017}. 

\begin{figure}
    \centering
    \includegraphics[width = \linewidth]{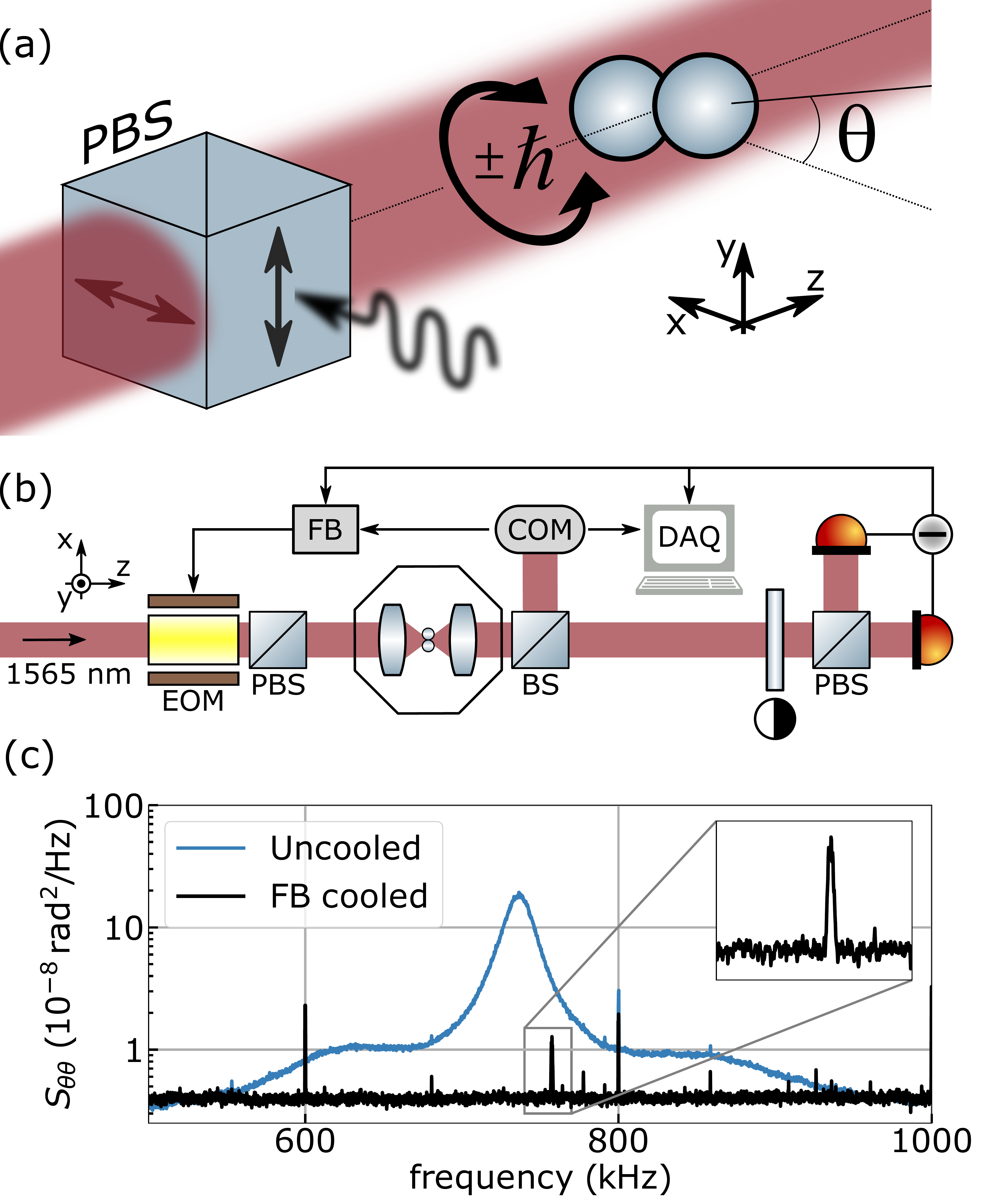}
    \caption{ (a)~Pictorial representation of radiation torque shot noise. An anisotropic scatterer in a linearly polarized light field experiences a fluctuating torque which arises from the vacuum fluctuations entering the unused port of the polarizing beamsplitter (PBS). (b)~Schematic of the experimental setup. Inside a vacuum chamber, we focus a laser beam (propagating along $z$, linearly polarized along $x$) with an aspheric lens ($0.7\,$NA) to form an optical trap. In the forward direction, the light is collected and split into two paths with a beamsplitter (BS). One half of the optical power is sent to a center-of-mass (COM) motion detector. The other half is used to measure the libration angle $\theta$ in a balanced detection scheme. The measurement is recorded with a data acquisition device (DAQ). The intensity of the laser beam [wavelength $\lambda = \SI{1565.0(1)}{\nm}$] is modulated with an electro-optic modulator (EOM) using feedback signals derived from the COM and the libration detector, respectively. 
    (c)~The blue line shows the measured power spectral density $S_{\theta\theta}$ of the libration motion at a pressure of $p_\mathrm{gas} = \SI{7.0(7)}{\milli\bar}$. The broad spectrum is a result of coupling between the angular degrees of freedom. The black line shows $S_{\theta\theta}$ at $p_\mathrm{gas} = \SI{1.1(1)E-8}{\milli\bar}$ and with feedback-cooling engaged for COM and librational motion, where the signal of the libration detector reduces to a single resonant line.}
    \label{fig:setup}
\end{figure}

Ideal testbeds for optomechanics with rotational degrees of freedom are optically levitated nanoparticles~\cite{Millen2020}. 
Control of their translational degrees of freedom has recently entered the quantum regime~\cite{Tebbenjohanns2020,Delic2020}. 
In a circularly polarized light field, such optically trapped particles can be spun at GHz rotation frequencies~\cite{Reimann2018,Ahn2018,Ahn2020}. 
In a linearly polarized field, an anisotropic particle aligns to the polarization direction, making this system an optically levitated librator~\cite{Kuhn2017control,Bang2020}. 
Importantly, the light field measures the angular orientation of the particle, and thus must give rise to measurement backaction in the form of radiation torque shot noise~\cite{Zhong2017,Seberson2020}. 
%In analogy to the case of translational motion, this torque noise can be pictured as a result of the particle's dipole moment, induced by the linearly polarized field, interacting with the vacuum fluctuations present in the orthogonal polarization direction. 
This torque shot noise is a result of the interaction between the dipole moment of the particle induced by the linearly polarized field, and vacuum fluctuations in the orthogonal polarization direction, as illustrated in Fig.~\ref{fig:setup}(a).
Alternatively, in a particle picture of light, the linearly polarized field scattering off the particle can be thought of as a stream of statistically independent left- and right-circularly polarized photons, each carrying angular momentum $\hbar$~\cite{Beth1936}.
Despite its fundamental importance, the observation of radiation torque shot noise has remained elusive. 

In this Letter, we report the observation of radiation torque shot noise driving the libration mode of an optically levitated rotor.
We trap a dumbbell-shaped dielectric nanoparticle in a linearly polarized laser beam, feedback-cool its librational motion, and investigate its reheating dynamics when cooling is switched off.
In high vacuum, we enter a regime where the reheating rate is independent of gas pressure. 
Our measurements reveal that in our system the radiation torque shot noise dominates over the torque noise of the thermal bath by more than a factor of four.
\paragraph{Experimental system.---} \label{sec:setup}
Our experimental setup is shown in Fig.~\ref{fig:setup}(b). 
We trap a dumbbell (composed of two silica spheres, nominal diameter $\SI{136}{\nano\meter}$) in a strongly focused laser beam [focal power $P = \SI{1050(50)}{\milli\W}$]. 
The beam propagates along the $z$ axis, and is linearly polarized along the $x$ axis. 
The laser power in the optical trap can be controlled with an electro-optical modulator. 
In the forward direction, the light from the trap is collected with a lens and divided at a beamsplitter. 
Half of the signal is sent to a center-of-mass (COM) motion detector~\cite{Gieseler2012}. 
The other half is sent through a polarizing beamsplitter and onto a balanced photodiode to detect the angular orientation of the dumbbell~\cite{Ahn2018,Reimann2018,vanderLaan2020}. 
%This orientation is described by two Euler angles
For small deviation angles of the dumbbell relative to the polarization axis, our detection scheme is sensitive to the angle $\theta$ of the dumbbell relative to the $x$ axis in the focal $xy$ plane~\cite{Seberson2019}.
Furthermore, the restoring torque generated by the light field on the dumbbell is to first order linear in $\theta$. 
The dumbbell is thus a harmonic oscillator with a libration frequency $\Omega_l$, following the equation of motion
\begin{equation}
    I\ddot\theta + I\gamma\dot\theta + I\Omega_l^2\theta = \tau_\mathrm{fl},
\end{equation}
with $I$ the moment of inertia of the dumbbell, $\gamma$ the damping rate, and each dot indicating a time derivative.
The fluctuating torque $\tau_\mathrm{fl}$ drives the librator. In this work, we demonstrate that at low pressures $\tau_\mathrm{fl}$ is dominated by the shot noise fluctuations of the light field.

The measured power spectral density $S_{\theta\theta}$ of the orientation angle $\theta$ at a pressure $p_\mathrm{gas} = \SI{7.0(7)}{\milli\bar}$ at room temperature is shown in Fig.~\ref{fig:setup}(c) in blue. 
The spectrum resembles a resonant line-shape, centered at $\SI{750}{\kilo\Hz}$, flanked by two broad shoulders on either side. 
This spectral shape has been explained as a consequence of the intricate rotational dynamics of the dumbbell, where the thermally driven spinning degree of freedom around the long axis of the dumbbell gives rise to an interaction between the two other orientational degrees of freedom~\cite{Seberson2019,Bang2020}.
We calibrate our detector signal by transforming the spectrum for $\theta$ to $\dot\theta$ and exploiting the equipartition theorem, according to the procedure detailed in Ref.~\cite{Hebestreit2018Calibration}.

At pressures $p_\mathrm{gas} < \SI{E-4}{\milli\bar}$, the gas damping is sufficiently low to apply effective feedback cooling to the libration and the center-of-mass motion. 
For both types of motion, we use the parametric feedback-cooling scheme originally developed for COM cooling~\cite{Jain2016} and suggested for libration cooling~ \cite{Seberson2019}. 
In this cooling technique, a phase-locked loop tracks the detector signal to generate a feedback signal at twice the oscillation frequency of the measured degree of freedom. 
This feedback signal is applied to the modulator controlling the power of the trapping laser, to generate a periodic modulation of the optical potential. 
A spectrum $S_{\theta\theta}$ under feedback cooling at $p_\mathrm{gas} =\SI{1.1(1)E-8}{\milli\bar}$ is shown in Fig.~\ref{fig:setup}(c) in black. 
Under feedback-cooling, the spectrum of the libration reduces to a single line centered at $\Omega_l=2\pi\times757~$kHz. 
The observed linewidth is limited by drifts of $\Omega_l$, arising from slow drifts of the laser power.
The area under the peak is a measure for the energy of the librator, and we extract a value of $E=\SI{0.24(3)}{\kelvin}$. Note that throughout this work, we normalize all energies by the Boltzmann constant to have the unit Kelvin.
This energy is a result of the balance of damping $\gamma$ and heating by the fluctuating torque $\tau_\mathrm{fl}$ acting on the librator.

\begin{figure}[b]
    \centering
    \includegraphics[width = \linewidth]{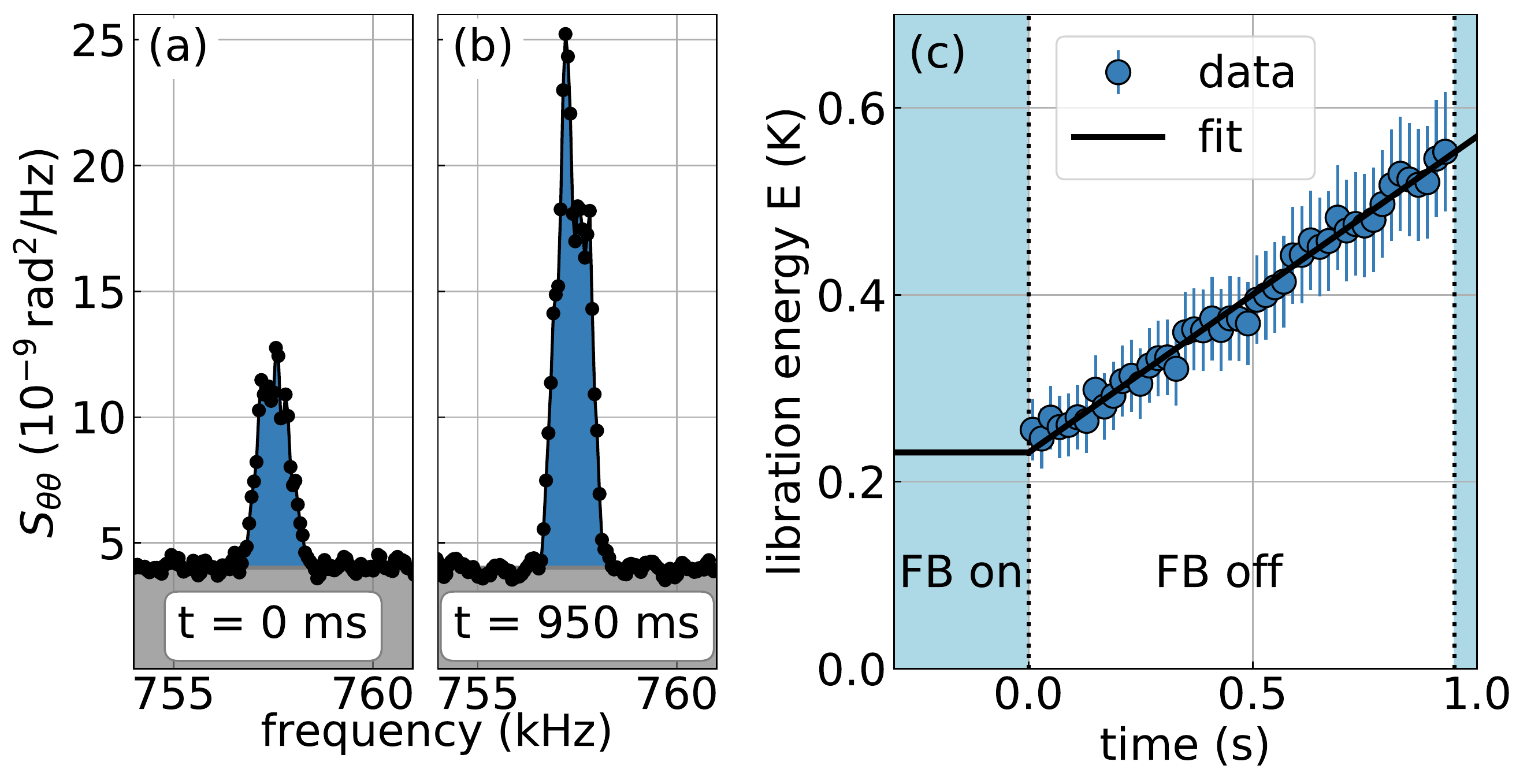}
    \caption{Reheating experiment at $p_\mathrm{gas} = \SI{1.1(1)E-8}{\milli\bar}$. (a)~Cooled libration spectrum right after feedback cooling is turned off. (b)~Libration spectrum after $\SI{950}{\milli \s}$, just before the feedback cooling is turned back on. (c)~Libration energy (blue circles) as a function of time. A linear fit to the data is shown as the solid line.}
    \label{fig:lib_spectrum_heating}
\end{figure}

\paragraph{Reheating protocol.---} \label{sec:experiment}
To quantify the torque fluctuations driving the levitated librator, we perform reheating experiments~\cite{Tebbenjohanns2019colddamping}. 
Each measurement cycle starts with the librator under feedback cooling. 
At time $t = \SI{0}{}$, we turn off the feedback for the libration (while center-of-mass cooling remains engaged) and measure the energy in the libration mode (extracted from the spectrum $S_{\theta\theta}$) as a function of time.
The cycle repeats as we re-engage feedback cooling of the libration. 
Since each experimental run records one realization of the stochastic reheating process, we repeat the cycle 400 times. 
In Fig.~\ref{fig:lib_spectrum_heating}(a), we show $S_{\theta\theta}$ averaged over all cycles at $p_\mathrm{gas}=\SI{1.1(1)E-8}{\milli\bar}$ at the beginning ($t = \SI{0}{\milli \s}$) of the reheating period, and in (b) at the end ($t = \SI{950}{\milli \s}$). 
%The spectra are broadened due to slow drifts in laser power over the cycles. 
We extract the energy of the librator by integrating the power spectrum (indicated by the blue shaded area), after subtracting the noise floor (grey area). 
The resulting mean libration energy is shown as a function of time in Fig.~\ref{fig:lib_spectrum_heating}(c). 
The heating process for the mean energy $E$ follows the equation
\begin{equation}
    E(t) = E_0 + (E_\infty-E_0)(1-\text{e}^{-\gamma t}),
\end{equation}
with $E_0=E(t=0)$, and $E_\infty$ the energy the system is equilibrating to. 
On a short timescale $t\ll\gamma^{-1}$, and for $E_\infty\gg E_0$, we find $E(t)=\gamma E_\infty t$. 
Thus, the slope of a linear fit to the data in Fig.~\ref{fig:lib_spectrum_heating}(c) yields the heating rate $\Gamma=\gamma E_\infty$.
\paragraph{Results and discussion.---} \label{sec:pressure}
\begin{figure}[t]
    \centering
    \includegraphics[width = 0.8\linewidth]{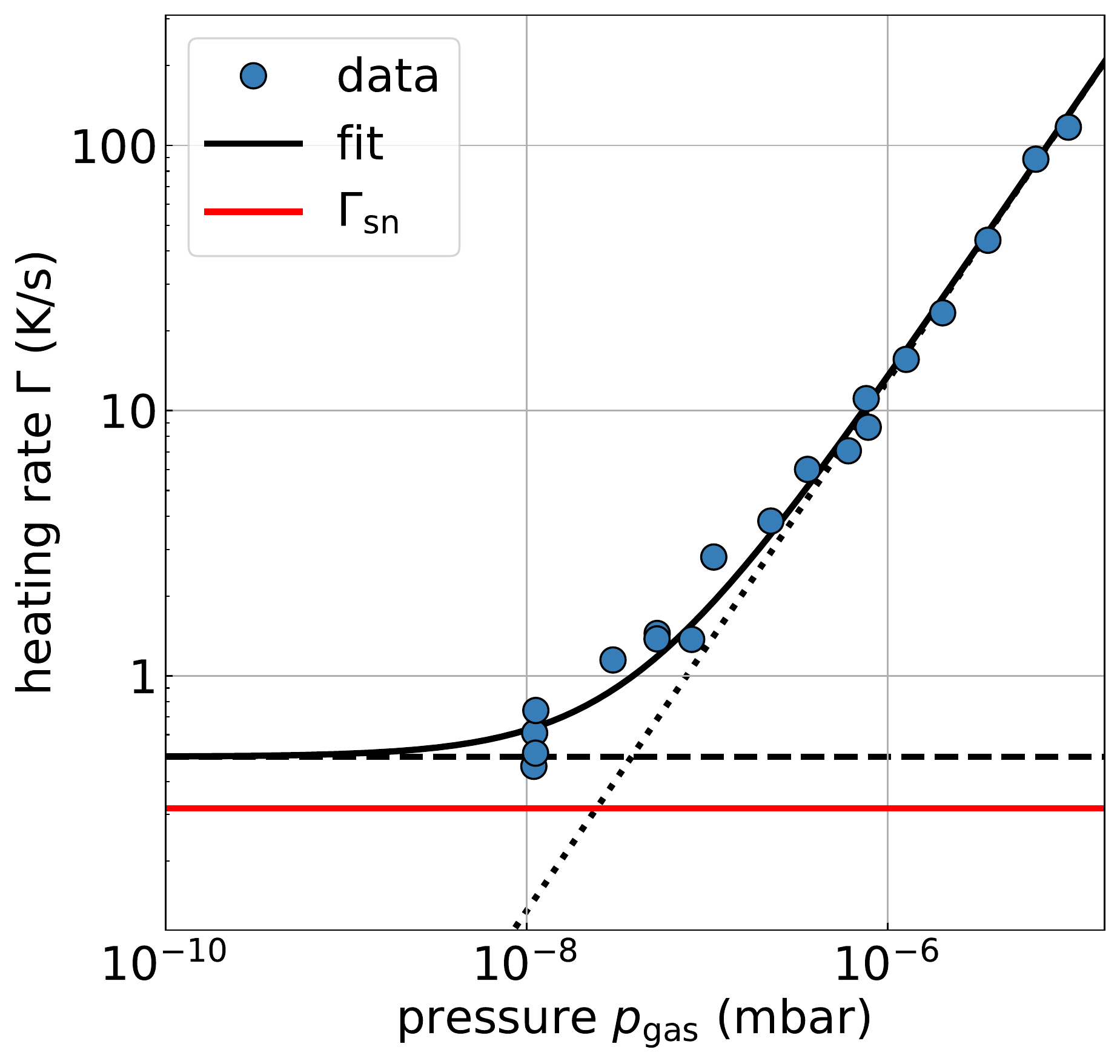}
    \caption{Heating rate (blue circles) as a function of pressure. The solid black curve shows a linear fit with constant offset $\Gamma_\mathrm{res}$ (dashed line). The dotted line indicates the contribution of the gas to the heating rate. The red line shows the theoretical prediction for the radiation torque shot noise heating rate $\Gamma_\mathrm{sn}$.}
    \label{fig:heating_pressure}
\end{figure}
Having established our protocol to measure the reheating rate $\Gamma$ of the levitated librator, we now investigate the origin of the fluctuating torque driving the reheating.
To quantify the contribution from the interaction with the residual gas in the vacuum chamber, we plot the measured reheating rate $\Gamma$ as a function of gas pressure in Fig.~\ref{fig:heating_pressure} as blue data points. 
At pressures above $\SI{E-7}{\milli\bar}$, the reheating rate scales linearly with pressure, as indicated by the dotted line. 
This scaling is expected, since the fluctuating torque due to the gas scales linearly with pressure.
At pressures below $\SI{E-7}{\milli\bar}$, we observe a significant deviation of the observed reheating rate from the linear scaling, and $\Gamma$ approaches a constant value. 
We fit our data with the function $\Gamma = a\times p_\mathrm{gas} + \Gamma_\mathrm{res}$, shown as the solid black curve in Fig.~\ref{fig:heating_pressure}, with the proportionality constant $a$ and the residual heating rate $\Gamma_\mathrm{res}$ as fit parameters. We obtain $\Gamma_\mathrm{res}=\SI{0.50(6)}{\kelvin\per\second}$.

The heating rate $\Gamma_\mathrm{sn}$ expected due to radiation torque shot noise is given by~\cite{Zhong2017, Ahn2018, Seberson2020}
\begin{equation}
    \Gamma_{\rm sn} = \frac{1}{2}\left(\frac{\Delta\alpha}{\alpha_x}\right)^2\hbar^2\frac{P}{I\hbar\omega},
\end{equation}
with $P$ the power scattered by the dumbbell, $\alpha_x$ its polarizability along the long axis, and $\Delta\alpha$ the difference in polarizability of long and short axes. 
For our experimental parameters~\footnote{We assume a scattered power of $\SI{80}{\micro\watt}$~\cite{Jain2016}, a ratio $\Delta\alpha/\alpha_x=0.14$~\cite{Ahn2018}, and a moment of inertia $I=\SI{1.6E-32}{\kg\m^2}$ measured by the method outlined in Ref.~\cite{vanderLaan2020}.}, we obtain $\Gamma_\mathrm{sn}= \SI{0.31}{\K \per \second}$, shown in Fig.~\ref{fig:heating_pressure} as the solid red line. This theoretical result is in good agreement with our measured $\Gamma_\mathrm{res}$. The difference between $\Gamma_\mathrm{res}$ and $\Gamma_\mathrm{sn}$ can be explained by the uncertainties of the refractive index, the dimensions of the dumbbell, the laser intensity in the trap, and the calibration procedure. 
We conclude that at a pressure $p=\SI{1.1(1)E-8}{\milli\bar}$ the radiation torque shot noise acting on the dumbbell exceeds the thermal torque noise by a factor of four. 
\begin{figure}[t]
    \centering
    \includegraphics[width = \linewidth]{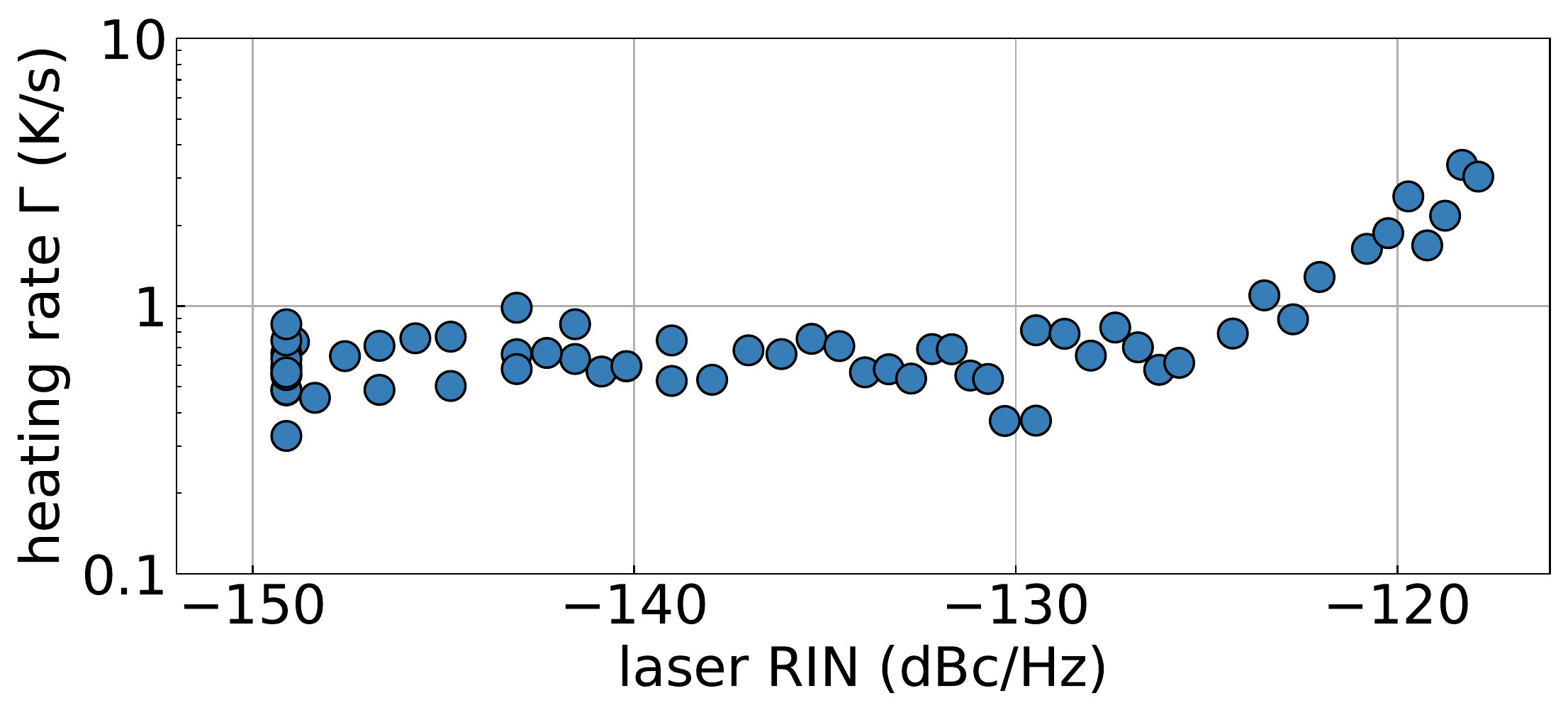}
    \caption{Heating rate as a function of RIN at \SI{1.1(1)E-8}{\milli\bar} (blue circles). An effect on the heating rate is observable only for RIN values exceeding \SI{-125}{\decibel c \per \Hz}. 
    }
    \label{fig:heating_RIN_main}
\end{figure}

Finally, we exclude classical laser noise as a source of the observed heating rate at low pressures. 
To this end, we introduce additional relative intensity noise (RIN) into the system by adding white noise with a bandwidth of $\SI{80}{\mega\Hz}$ and variable variance to the feedback signal entering the electro-optical modulator.
Without added noise, our laser has a measured RIN of $\SI{-149}{\decibel c \per \Hz}$ (at both $\Omega_\mathrm{l}$ and $2\Omega_\mathrm{l}$). In Fig.~\ref{fig:heating_RIN_main}, we plot the heating rate measured at a pressure of \SI{1.1(1)E-8}{\milli\bar} as a function of laser RIN (blue circles).  
The heating rate remains constant up to a RIN of $\SI{-125}{\decibel c \per \Hz}$ and increases only for higher values of RIN. 
We therefore conclude that the influence of the baseline RIN on the heating rates reported in Fig.~\ref{fig:heating_pressure} is negligible.

\paragraph{Conclusion.---} \label{sec:conclusion}

We have observed the effect of radiation torque shot noise on a mechanical rotor for the first time. 
In particular, we have demonstrated that this torque noise dominates the heating rate of the libration mode of a dumbbell trapped in a linearly polarized laser beam in high vacuum. 
Our work is of significance for the development of torque sensors based on levitated nanoparticles~\cite{Ahn2020}, with potential applications for the characterization of materials at the nanoscale~\cite{Wu2017,Kim2017,Losby2018}, and for the detection of angular momentum states of light~\cite{Andrews2012}. Our experiments constitute an important step towards operating those sensors at the standard quantum limit, which requires careful balancing of measurement backaction with intrinsic damping~\cite{Kim2016}. 
At this limit, levitated torque sensors hold promise to provide access to currently elusive but deeply fundamental effects of vacuum friction~\cite{Kardar1999,Manjavacas2010,Zhao2012,Manjavacas2017}.
Furthermore, entering the backaction-limited regime is a necessary requirement to achieve quantum control over optomechanical systems~\cite{Rossi2018}, with the aim to test quantum mechanical effects in rotating systems at a macroscopic scale~\cite{Stickler2018,Stickler2018rotationquantum}.
Importantly, we establish parametric feedback-cooling as a powerful technique to control rotational motion. 
Therefore, this work brings ground-state cooling and quantum control of optically levitated librators firmly within reach.

\bibliographystyle{apsrev4-1}
\bibliography{bibliography}

\end{document}